\documentclass[nofootinbib,prd,aps,superscriptaddress,preprintnumbers,preprint]{revtex4-1}\usepackage{epsf}
\pdfoutput=1
\usepackage[english]{babel}
\usepackage{amsmath,amssymb,bm,slashed}
\usepackage{multirow}
\usepackage{graphicx}
\usepackage[sort&compress]{natbib}
\usepackage{xcolor}
\usepackage[normalem]{ulem}

\definecolor{red}{rgb}{1.0, 0, 0}

\allowdisplaybreaks

\bibpunct{[}{]}{,}{n}{}{,}

\newcommand{\tev}{\, {\rm TeV}}
\newcommand{\gev}{\, {\rm GeV}}

\newcommand{\beq}{\begin{equation}}
\newcommand{\eeq}{\end{equation}}
\newcommand{\be}{\begin{equation}}
\newcommand{\ee}{\end{equation}}
\newcommand{\ba}{\begin{array}}
\newcommand{\ea}{\end{array}}
\newcommand{\beqa}{\begin{eqnarray}}
\newcommand{\eeqa}{\end{eqnarray}}
\newcommand{\bea}{\begin{eqnarray}}
\newcommand{\eea}{\end{eqnarray}}
\newcommand{\beqn}{\begin{eqnarray}}
\newcommand{\eeqn}{\end{eqnarray}}

\newcommand{\lsim}{\stackrel{<}{_\sim}}
\newcommand{\gsim}{\stackrel{>}{_\sim}}

\definecolor{red}{cmyk}{0,1,1,0.4}

\def\eq#1{eq.~(\ref{#1})}
\def\EmissT{\,{}/ \hspace{-8pt}  E_{T}}

\begin{document}
{\flushleft{\vspace*{-1cm}\tiny{CERN-PH-TH/2013-037 }}}

\def\Cincy{Department of Physics, University of Cincinnati, Cincinnati, Ohio 45221,USA}
\def\CERN{CERN Theory Division, CH-1211 Geneva 23, Switzerland}
\def\Weiz{Department of Particle Physics and Astrophysics, Weizmann Institute of Science, Rehovot 76100, Israel}

\title{Flavoured Naturalness}
\author{Monika \surname{Blanke}}            
\affiliation{\CERN}
\author{Gian F.\ \surname{Giudice}}            
\affiliation{\CERN}
\author{Paride~\surname{Paradisi}}           
\affiliation{\CERN}
\author{Gilad \surname{Perez}}            
\affiliation{\CERN}
\affiliation{\Weiz}
\author{Jure \surname{Zupan}}              
\affiliation{\Cincy}
\date{February 28, 2013} 

\begin{abstract}
We show that a large mixing between the right-handed charm and top squarks {\it (i)} is allowed by low-energy flavour constraints;  {\it (ii)} reduces the experimental bound on the stop mass; {\it (iii)} has a mild, but beneficial, effect on fine-tuning; {\it (iv)} leads to interesting signatures at the LHC not presently investigated by experiments. We estimate the current bound on the stop mass, in presence of flavour mixing, and discuss the new collider signatures. The signal in the $t\bar c(c\bar t)+\EmissT$
channel is large enough that it can be immediately searched for experimentally, while the signature with same-sign tops and $\EmissT$ requires a luminosity upgrade of the LHC.
\end{abstract}

\maketitle

\section{Introduction}

Now that the Higgs boson has been discovered~\cite{:2012gk,:2012gu}, naturalness becomes the most pressing question  within reach of the LHC. 
The spectre of 
unnaturalness, after making its first apparition with the cosmological constant, is now looming behind the Higgs boson, as no new physics has been sighted at the LHC. Casting the final verdict on naturalness is one of the main tasks of the high-energy run at 14 TeV.

Since the main contribution that destabilises the electroweak scale in the Standard Model (SM) comes from a top-quark loop, the key for addressing the naturalness problem at the LHC is the identification of the top partner that is responsible for cancelling the quadratic divergence. The top partner can be a scalar particle (such as the stop, in the case of supersymmetry) or a fermion (as in the case of composite Higgs models). The hunt for the top partner, which is actively pursued at the LHC, is the most direct way to test the idea of naturalness for the electroweak scale and therefore it is of the utmost importance. As the search has been so far unsuccessful, it is mandatory to explore all possible variations of the theory, especially those in which the top partner could have escaped detection. Such studies could uncover new experimental signals different than those currently pursued. In this paper we explore one such possibility: the effect of flavour mixing between the top partner and the other states (for previous related studies, see refs.~\cite{Han:2003qe,Liu:2004bb,Eilam:2006rb,Cao:2006xb,Cao:2007dk,LopezVal:2007rc,Bejar:2008ub,Hiller:2008wp,Kribs:2009zy,Bartl:2010du,Bartl:2012tx,Herrmann:2011xe,Bruhnke:2010rh,Hurth:2009ke}). 

Let us focus on the case of supersymmetry. The leading-log one-loop contribution to the Higgs mass parameter $m_{H_u}^2$ is
\beq
\delta m_{H_u}^2 =  - \frac{\beta_{H_u}}{16\pi^2}\ln \frac{\Lambda}{m_{\tilde t}}, 
\eeq
with
\beq
\beta_{H_u} = 6\, {\rm Tr} \left( Y_u^\dagger {\tilde m}_Q^2 Y_u + Y_u {\tilde m}_U^2 Y_u^\dagger +A_u^\dagger A_u \right) + \cdots \, .
\label{beta}
\eeq
Here $\Lambda$ is the scale of supersymmetry-breaking mediation, $Y_u$ is the up-type Yukawa matrix, ${\tilde m}_{Q,U}$ are the left and right up-squark mass matrices, and $A_u$ is the matrix of trilinear couplings. The trace in \eq{beta} is taken over flavour indices. Since $m_{H_u}^2$ sets the scale of electroweak breaking, the size of the mass parameters in $\beta_{H_u}$ gives a measure of the tuning in the theory.

To a good approximation in the up quark mass basis, we can write $Y_u = {\rm diag}(0,0,y_t)$, where $y_t$ is the top Yukawa coupling, and therefore the matrix $Y_u$ in \eq{beta} effectively acts as a projector of the squark mass matrices onto their (3,3) elements in the supersymmetric up mass basis. In the absence of flavour mixing, these elements correspond to the stop mass parameters. In the general case of flavour mixing, they are not equal to the stop mass parameters, but instead give the absolute upper bound on the smallest eigenvalue (as follows from the hermiticity of the squark square-mass matrices). From this result, one would naively conclude that having no flavour mixing is the most favourable situation to minimise the amount of fine tuning, while satisfying the experimental bounds on stop masses. Indeed, in the presence of flavour mixing, the mass parameters that enter $\beta_{H_u}$ are necessarily larger than the physical stop masses.

However, this conclusion is incorrect because the experimental limits on squark masses depend in a non-trivial way on quark flavour and mixing angles. In particular, the LHC constrains the charm squark more poorly than the stop -- because of the absence of top and/or bottom quarks in the final state,  and also more poorly than the up-squark -- because of the smaller cross section due to the limited charm content of the proton and reduction of efficiency of searches when looking at lower squark masses~\cite{Mahbubani:2012qq}. A mixing between $\tilde t$ and $\tilde c$ can then lead to weaker experimental bounds on the stop (or, more correctly, on the mass eigenstate which is predominantly stop) and a reduction of the size of the mass combination entering $\beta_{H_u}$. 
In this context let us stress that with a bigger data sample from the upcoming $14\tev$ run the difference between the stop and scharm mass bounds are expected to further increase, the reason being the much smaller SM background for the $t\bar t+\EmissT$ final state with respect to $\text{jets}+\EmissT$.

For example, in the case of mixing between $\tilde t_R$ and $\tilde c_R$ (neglecting left-right mixing and flavour mixings with the first generation as well as the contribution from the left-handed stop), we find  $\beta_{H_u}\sim 6\,(c^2\, m_1^2+s^2\, m_2^2)$, where $s$ and $c$ denote sine and cosine of the $\tilde t_R$--$\tilde c_R$ mixing angle, and $m_{1,2}$ are the two squark mass eigenvalues.\footnote{Throughout this paper we use $s$ and $c$ as a short-hand notation for $\sin \theta^{ct}_R$ and $\cos \theta^{ct}_R$, where $\theta^{ct}_R$ is the mixing angle between $\tilde t_R$ and $\tilde c_R$. We also take $\tilde q_1$ to be the stop-like state in the limit of small mixing $s\to 0$.} In this case, the characteristic stop signature with two tops and $\EmissT$ has a rate given by $c^4\, \sigma(m_1) + s^4\, \sigma(m_2)$, where $\sigma(m_{1,2})$ are the production cross-sections for the two squarks, respectively. While this combination is constrained by the negative stop searches, only weaker limits apply to the $\tilde c$ production signal with no tops and $\EmissT$, whose rate is given by $s^4\, \sigma(m_1) + c^4\,\sigma(m_2)$. The novelty of the flavour-mixed case is that a new channel with one top and $\EmissT$ is now present, with rate given by $2\, s^2\, c^2\,(\sigma(m_1) +\sigma(m_2))$. 
This process has previously been discussed in refs.~\cite{Bartl:2010du,Bartl:2012tx,Bruhnke:2010rh}. In sect.~\ref{sec:constraints} we study these signals, finding three interesting results. {\it (i)} The experimental limit on the mostly-stop state is weaker than the corresponding limit for a pure stop state. {\it (ii)} Our best estimate of the combination of the limits from all three channels allows for the existence of scharm--stop mixed states that give a smaller contribution to $\beta_{H_u}$ (and therefore to the amount of tuning) than the case of pure states. {\it (iii)} The new channel with a single top and $\EmissT$ gives the most unambiguous way to detect the stop flavour-mixing angle. 

We certainly do not claim that flavour mixing can remove the tuning problem, nowadays ubiquitous in supersymmetric theories. 
The contribution to $\beta_{H_u}$ that we have studied is not the only source of tuning, even though it is the most straightforward one to test experimentally. An additional important contribution comes from a large $A_t$ coupling, which is needed to account for the Higgs boson mass in minimal models with relatively small stop masses. 
Nevertheless, it is interesting that a flavour mixing in the stop sector not only is harmless for the tuning, but it can even be beneficial. 
In our view, this strengthens the motivation for a thorough experimental exploration of non-trivial flavour structures involving the stops. 
We should also mention that even without a large $A_t$, stops can be lighter in extensions of minimal supersymmetry in which the Higgs mass receives contributions from other sources (see {\it{e.g.}}~\cite{Hall:2011aa} for a recent discussion and Refs. therein) leading to a lower level of fine tuning that is identified with the contributions of the top partners in Eq.~(\ref{beta}).

Before discussing the collider signatures of flavour mixing in the stop sector, it is necessary to check if large mixing angles are allowed by low-energy processes. Flavour violations in the left-handed squarks are constrained because their effect in the up and down sectors are related by a CKM rotation. If the structure of flavour violation is generic, constraints from $B_s^0$--${\bar B}_s^0$ mixing and $B\to X_s \gamma$ set significant limits on the $\tilde t$--$\tilde c$ mixing angle (see {\it e.g.}~\cite{Giudice:2008uk} for more details). 
On the other hand, if the quark and squark mass matrices are simultaneously diagonal in the down sector, the constraints on mass splittings between $\tilde t$ and $\tilde c$ become fairly weak (see~\cite{Gedalia:2010mf}  for a recent discussion on this limit), but their mixing angle is small because it is determined by the CKM matrix. In conclusion, flavour constraints severely limit the case of a large $\tilde t$--$\tilde c$ mixing angle in the left sector. A possible way out is to take gluinos in the multi-TeV region and small $\tan\beta$, such that flavour limits from chargino exchange are relatively mild. 

For simplicity, we prefer to concentrate on the case of flavour violation in the right-handed sector. The main constraint comes from $D^0$--${\bar D}^0$ mixing, which sets bounds on up-type squark mixing angles.
Assuming large splitting in the squark sector one can constrain the up-charm squark mixing angle  $\theta^{uc}_R<0.05\, (\tilde{m}/500~{\rm GeV})$, for comparable squark and gluino masses~\cite{Gedalia:2009kh}, and the product of the up-top and charm-top mixing $\theta^{ut}_R\, \theta^{ct}_R<0.01\, (\tilde{m}/500~{\rm GeV})$, in the case of light stop and decoupled first two generation squarks~\cite{Giudice:2008uk}. In the limit of no admixture with the first generation squarks, the mixing angle between $\tilde t_R$ and $\tilde c_R$ is left unconstrained. Such mixing can induce the decays $t\to c Z$ and $t\to ch$ at one loop and, as discussed below, can lead to same-sign top production~\cite{Bartl:2010du}.  However, the experimental constraints on all these processes are too weak to be significant. The $\tilde t_R$--$\tilde c_R$ mixing can induce flavour violation in the down sector through higgsino loops, but the effect is always proportional to charm Yukawa couplings, and therefore negligible. In conclusion, the mixing angle $\tilde t_R$--$\tilde c_R$ could be large, even maximal, without any conflict with present flavour constraints, also when squarks are relatively light.

This situation leads to characteristic signatures at the LHC and deserves experimental attention. One additional interesting aspect of this scenario is that $\tilde t \to t \chi_1^0$, where $\chi_1^0$ is a gaugino, produces a polarized top~\cite{Perelstein:2008zt} (see also~\cite{Gedalia:2009ym,Belanger:2012tm}). The  semi-leptonic decays of right-handed tops lead to a harder lepton spectrum~\cite{Almeida:2008tp} and smaller lepton-$b$ angular separation~\cite{Rehermann:2010vq} compared to the decays of left-handed tops. As a result, one expects different sensitivities of the searches for the purely right-handed and purely left-handed stops (decaying to top and neutralino). This is in accordance with  the recent experimental analyses~\cite{CMStalk}, which give a stronger bound on right-handed stops. Therefore, right-handed stops give a leading source of pressure on naturalness.
In the limit of heavy higgsinos, only the left-handed stop can decay to a chargino and bottom quark: the  resulting bound strongly depends on the mass splitting between the chargino and the lightest supersymmetric particle and is therefore model-dependent.

\section{Stop-scharm mixing at the LHC}
\label{sec:constraints}

\subsection{Current constraints}

In order to estimate the bounds on the masses $ m_1, m_2$ and the mixing parameter $c=\cos\theta^{ct}_R$ of the mixed $\tilde t_R -\tilde c_R$ states, we assume 100\% branching ratios for the decays of the flavour eigenstates $\tilde t_R \to t\chi_1^0$ and $\tilde c_R \to c\chi_1^0$, as well as a massless and purely gaugino lightest supersymmetric particle (LSP). {While it is straightforward to extend our analysis to non-zero LSP masses, we content ourselves here with assessing its impact  in a qualitative manner. For small LSP masses $\lsim 150\gev$ the experimental bounds on the squark masses are almost independent of the value of the LSP mass, hence in this case we expect a very small impact on our analysis. For larger LSP masses the on average smaller $p_T$ of the decay products becomes relevant, resulting in weaker constraints on the squark masses. The most important effect of finite LSP mass is the difference in phase space between the $t\chi_1^0$ and $c \chi_1^0$ final states, which becomes relevant if $m_t + m_{\chi_1^0}\lsim m_{1,2}$. This leads to a suppression of the $t\chi_1^0$ final state with respect to the $c\chi_1^0$ one. Consequently since the bound on the $\text{jets}+\EmissT$ final state is weaker than the one on $t\bar t+\EmissT$, the phase space suppression is beneficial for lowering the mass bounds on the mixed $\tilde c_R -\tilde t_R$ states.}

The two squark states are pair produced at the LHC by strong interactions. Since the charm quark PDF is very small, the main production mechanism is gluon fusion, and therefore the production cross section $\sigma$ is a function only of the squark mass $m_{1,2}$, and is independent of the flavour admixtures. Subsequently each squark decays into $t\chi_1^0$ and $c\chi_1^0$ with branching ratios determined by their flavour content -- {\it i.\,e.}  $BR_1(t\chi_1^0) = BR_2(c\chi_1^0) =c^2$ and $BR_1(c\chi_1^0)=BR_2(t\chi_1^0) =s^2$. We neglect the phase space suppression of the $t\chi_1^0$ final state relative to the $c\chi_1^0$ final state which would become relevant in the case of a compressed spectrum.

Direct pair production of the mixed $\tilde t_R -\tilde c_R$ states gives rise to the signatures $t\bar t + \EmissT$, $c\bar c + \EmissT$ and $t\bar c (c\bar t) + \EmissT$. While no dedicated search for the latter signature has been performed so far, both ATLAS and CMS have analysed the $t\bar t+\EmissT$ and jets+$\EmissT$ final states and placed bounds on the masses of the stop and the first two generation squarks, respectively.

\begin{figure}/,
\centering
\includegraphics[width=0.54\textwidth]{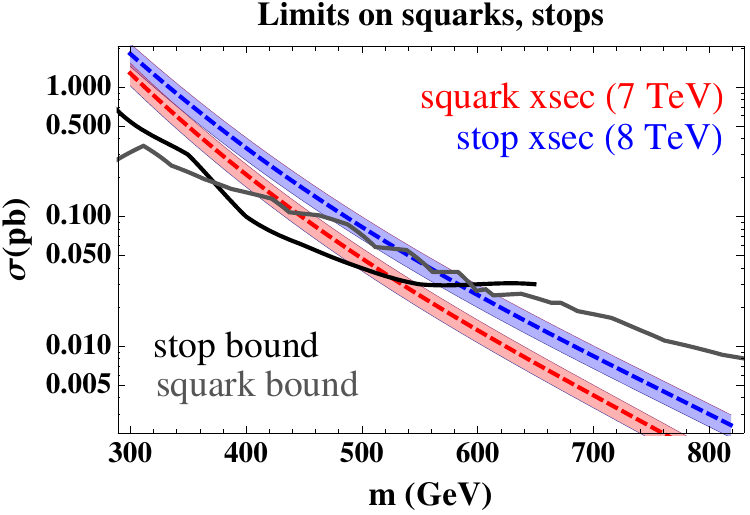}
\caption{\label{fig:xsecbounds} Various upper limits on squark pair-production cross-sections. In gray we show the envelope of experimental bounds on right-handed scharm for a massless neutralino, from $7\,{\rm TeV}$ $5\,{\rm fb}^{-1}$ ATLAS and CMS $\text{jets}+\EmissT$ searches~\cite{ATLAS:2012ona, :2012mfa, CMS:2012yua,Chatrchyan:2012wa, Aad:2012fqa}. In black we show the corresponding limit on the stops production taken from~\cite{ATLAS-CONF-2012-166}. The red (blue) band corresponds to the theory  prediction for the scharm (stop) pair production~\cite{Beenakker:2011fu}, the dashed line being the central value.}
\end{figure}

Concerning the $t\bar t + \EmissT$ final state, the strongest bounds have been obtained from the search for a single isolated lepton, jets, and large  $\EmissT$. While the CMS cross section limits \cite{CMS-PAS-SUS-12-023} are obtained for unpolarized top quarks,  the corresponding ATLAS analysis \cite{ATLAS-CONF-2012-166} assumes a simplified model where the tops are almost exclusively right-handed. Therefore the cross section limits from the latter analysis can directly be applied to our scenario of mixed $\tilde t_R -\tilde c_R$ states.
The jets+$\EmissT$ searches from ATLAS \cite{ATLAS:2012ona,Aad:2012fqa} and CMS \cite{:2012mfa,CMS:2012yua,Chatrchyan:2012wa}  constrain both the $c\bar c + \EmissT$ and $t\bar c (c\bar t) + \EmissT$ final states. The constraints on a single light second generation squark have been analysed in detail in \cite{Mahbubani:2012qq}, and that study applies directly (modulo flavour mixing) to the $c\bar c + \EmissT$ final state in our scenario. In Fig.~\ref{fig:xsecbounds} we show the various upper limits on squark pair-production cross-sections. In gray we show the envelope of 95\% CL experimental bounds on right-handed scharm for a massless neutralino \cite{Mahbubani:2012qq}, and in black we show the corresponding 95\% CL limit on the stop pair production \cite{ATLAS-CONF-2012-166}. The red (blue) band corresponds to the theory prediction for the scharm (stop) pair production at 7TeV (8TeV) at NLO+NLL {in the limit of decoupled gluino}~\cite{Beenakker:2011fu}.

 Obtaining sensible constraints on the $t\bar c (c\bar t) + \EmissT$ final state is much more involved, since no dedicated search for this final state has been performed yet. In order to obtain an estimate of how strongly the $\tilde t_R -\tilde c_R$ mixing scenario is constrained at present, we therefore employ two extreme approaches:
\begin{enumerate}
\item
{\it Conservative estimate:} We assume the $t\bar c (c\bar t) + \EmissT$ final state with a hadronically decaying top to contribute to the jets+$\EmissT$ signature with the same efficiency as the $c\bar c + \EmissT$ final state. In other words we treat the hadronic top like a normal jet. While this approximation is reasonable in the limit of very boosted tops, it clearly is too conservative if the tops are not very energetic.
\item
{\it Aggressive estimate:} We neglect the impact of the  $t\bar c (c\bar t) + \EmissT$ final state on the jets+$\EmissT$ searches and assume that there are no constraints available on this signature.
\end{enumerate}
We are aware that neither of the two estimates realistically accounts for the present situation. Nonetheless they give us a good indication of the presently available constraints, and we expect the true situation to lie in between these two extreme cases.

In order to constrain the masses and mixing angle of the mixed $\tilde t_R -\tilde c_R$ system, we construct a $\chi^2$ function
\beq
\chi^2=\left[\frac{c^4 \sigma(m_1)+r_{t\bar t} s^4 \sigma(m_2)}{\Delta \sigma_{t\bar t}(m_1)}\right]^2+\left[\frac{(s^4+ 2 s^2 c^2B_W) \sigma(m_1)+r_\text{jets} (c^4+ 2 s^2 c^2B_W) \sigma(m_2)}{\Delta \sigma_\text{jets}(m_1)}\right]^2,
\label{eq-chi2}
\eeq
where $\Delta \sigma_f(m_1)$ is the $1\sigma$ upper bound deduced from the measured 95\%CL bounds assuming Gaussian errors and zero mean ($\Delta \sigma_f\equiv\sigma_f^{95\%CL}/1.96$), and the factor $r_f\equiv \Delta \sigma_f(m_1)/\Delta \sigma_f(m_2)$ (for $f=t\bar t$, jets) is the ratio of experimental sensitivities to squarks with masses $m_1$ and $m_2$  and is a naive guess for how much the higher mass squarks contribute to the analysis. Note that Eq. \eqref{eq-chi2} has the right limit when $m_1\approx m_2$. 
We emphasize that here we only consider incoherent production processes. This is valid as long as
$\left|  m_1-m_2\right|\gg \Gamma_{1,2}$, where $ \Gamma_{1,2}$ stands for the width of the two states respectively. As the typical squark width-to-mass ratio is ${\mathcal O}(10^{-2})$, this is a very good approximation.
The above expression is also correct if $\Delta \sigma_f$ is $m_i$ independent ({\it i.e.}  $r_f=1$), as then these are just counting experiments and  the higher mass events are fully counted. In Eq. \eqref{eq-chi2} we give the higher mass events a higher weight $(r_f>1)$ to account for larger experimental sensitivity to the higher mass squarks. We assume Gaussian errors throughout so that Eq. \eqref{eq-chi2} has a $\chi^2$ distribution for two degrees of freedom. 
In the conservative scenario we set the branching ratio $B_W\equiv BR(W\to$ hadrons) equal to its SM value $B_W=2/3$, 
and in the aggressive approach we set $B_W=0$, so that the impact of the $t\bar c(c\bar t)+\EmissT$ final state on the $\text{jets}+\EmissT$ search is neglected.

\begin{figure}
\centering
\includegraphics[width=0.49\textwidth]{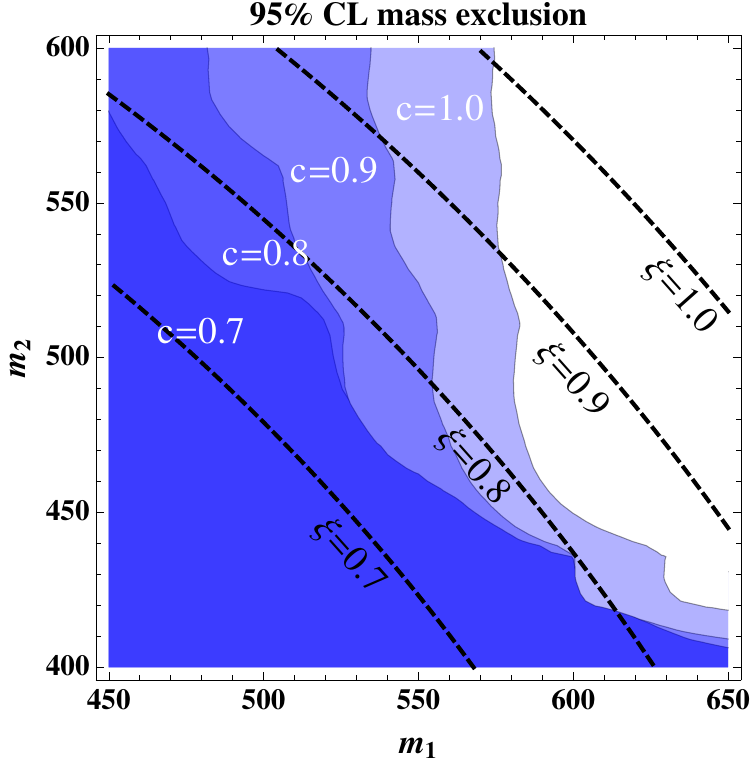}
\caption{\label{fig:m1m2constraints}Exclusions at 95\% CL for $m_1$ and $m_2$ (the masses of the mostly-top and mostly-charm squarks, respectively) from $t\bar t+\EmissT$ and from $\text{jets}+\EmissT$ in the conservative approach, fixing $c\equiv \cos\theta_R^{ct}=0.7,0.8,0.9,1.0$ (from darker to lighter shades, as indicated). Contours of constant tuning parameter $\xi$, obtained for $c = 0.7$, are displayed by the dashed lines.}
\end{figure}

In Fig.\ \ref{fig:m1m2constraints} we show the  95\% confidence level (CL) exclusion contours in the $m_1-m_2$ plane obtained in the conservative approach, for fixed values of the mixing angle. We observe that with increasing mixing angle the constraint on the stop-like state $m_1$ becomes significantly weaker, while the bound on the scharm-like state $m_2$ remains mostly unaffected. With nearly maximal mixing angle ($c\approx 0.7$), squark masses of about $ 520\gev$ are allowed for both states simultaneously.
{The wiggles in the exclusion contours are a direct consequence of the wiggles in the experimental upper bound on the scharm pair production cross section, see the grey line in Fig.\ \ref{fig:xsecbounds}.}

In order to quantify the improvement obtained for the fine-tuning in the Higgs mass parameter $\delta m_{H_u}^2$, we define the tuning parameter
\beq
\xi=\frac{c^2 m_1^2 + s^2 m_2^2}{m_0^2},
\eeq
where $m_0=585$ GeV is the experimental bound on the right-handed stop mass without mixing \cite{ATLAS-CONF-2012-166}. In Fig.\ \ref{fig:m1m2constraints} contours of constant $\xi$ are overlaid, in this case always setting  $c=0.7$ for simplicity. We see that values below $\xi = 0.8$ are allowed for large mixing, so that a marginal improvement is possible.

\begin{figure}
\includegraphics[width=0.49\textwidth]{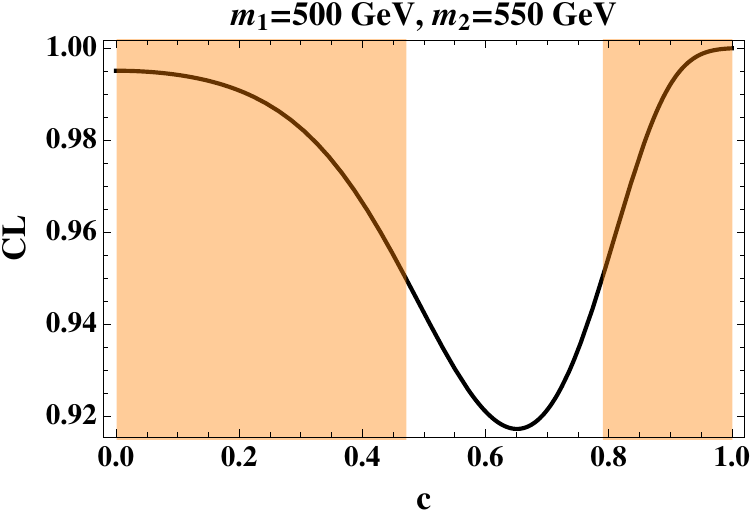}\hfill
\includegraphics[width=0.49\textwidth]{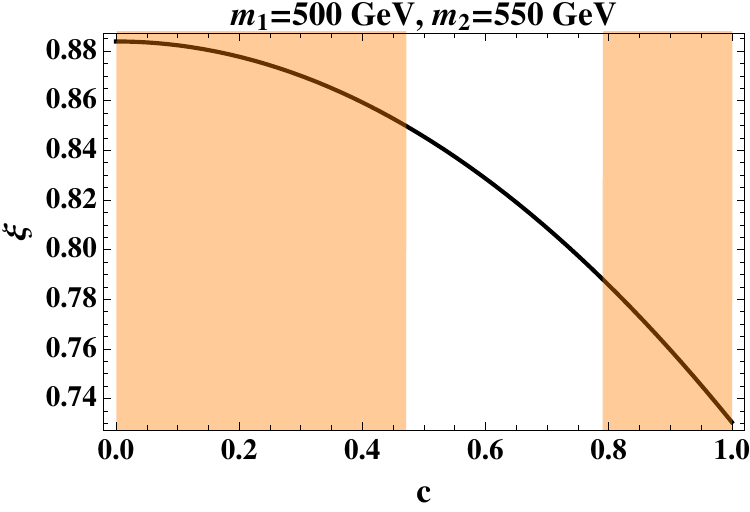}

\caption{\label{fig:Xi_CL_450_500}The confidence level (CL) of exclusion and the tuning parameter $\xi$ as functions of $c=\cos\theta^{ct}_R$ for the case $m_1=500\gev$ and $m_2=550\gev$. Shaded regions are excluded at 95\%CL.}
\end{figure}

Let us now fix the squark masses to $m_1=500\gev$ and $m_2=550\gev$ and study the effect of flavour mixing in more detail. To this end we show in Fig.\ \ref{fig:Xi_CL_450_500} the confidence level of exclusion and the tuning parameter $\xi$ as functions of the mixing parameter $c$. We observe that, for this choice of masses, a large range of mixing angles ($c \lsim 0.5$ and $c \gsim 0.8$) is excluded, but there still exists an interval around maximal mixing ($c\approx 0.7$) where the confidence level drops 
below 95\% CL and such low masses are allowed. In this window the fine-tuning parameter $\xi$ is around $0.8$.

\begin{figure}
\centering
\includegraphics[width=0.49\textwidth]{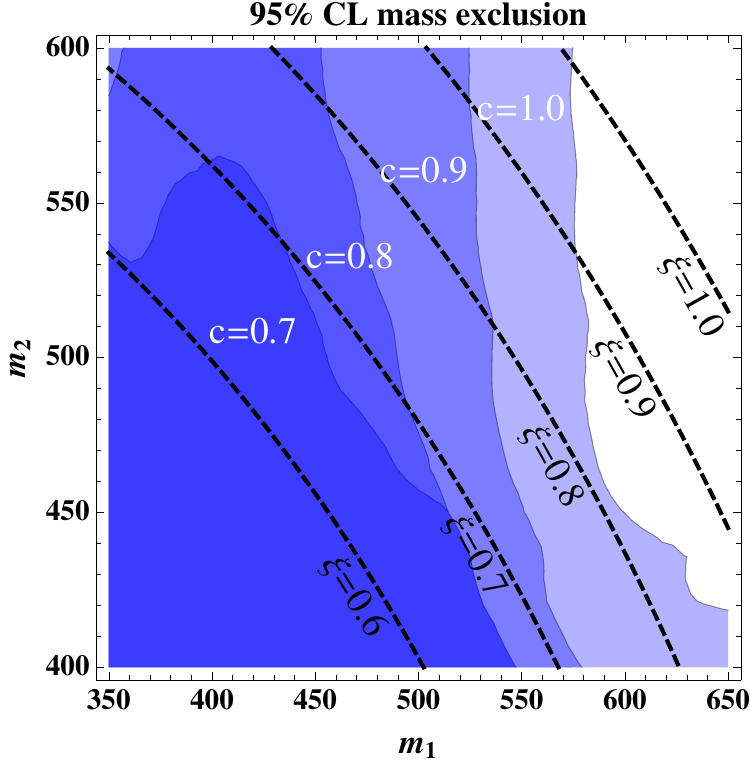}
\caption{\label{fig:m1m2constraints_alter}Exclusions at 95\% CL for $m_1$ and $m_2$ (the masses of the mostly-top and mostly-charm squarks, respectively) from $t\bar t+\EmissT$ and from $\text{jets}+\EmissT$ in the aggressive approach, fixing $c\equiv \cos\theta^{ct}_R=0.7,0.8,0.9,1.0$ (from darker to lighter shades, as indicated). Contours of constant tuning parameter $\xi$, obtained for $c = 0.7$, are displayed by the dashed lines.}
\end{figure}

Having seen that already the conservative approach leads to weaker bounds on squark masses, we now repeat our study for the aggressive approach.  The result is shown in Fig.~\ref{fig:m1m2constraints_alter} which is analogous to Fig.~\ref{fig:m1m2constraints} above. As expected, the constraints are weaker than before, whenever the mixing is large. This originates from the fact that we neglect now the constraint from the $t\bar c(c\bar t)+\EmissT$ final state, whose rate is largest for maximal mixing. Masses $\sim450-500\gev$ are now allowed given a large mixing angle and the tuning can be reduced by at most 40\%. 
The case where the mixing is nearly maximal leads to an interesting feature in the exclusion curve. 
For a fixed value of $m_2\sim 550\,$GeV the exclusion contour in $m_1$ is not single valued.  This stems from the fact that the bound on the stop cross section (shown by the black curve in Fig.~\ref{fig:xsecbounds}) is steeply falling near stop masses of $\sim 380\,$GeV, faster than the corresponding theory prediction for the stop pair production. This implies that unlike what one would naively expect, for a fixed $m_2$ value, when the mass of $m_1$ is increased the bound becomes stronger, for this narrow region of $m_1$.

\begin{figure}
\includegraphics[width=0.49\textwidth]{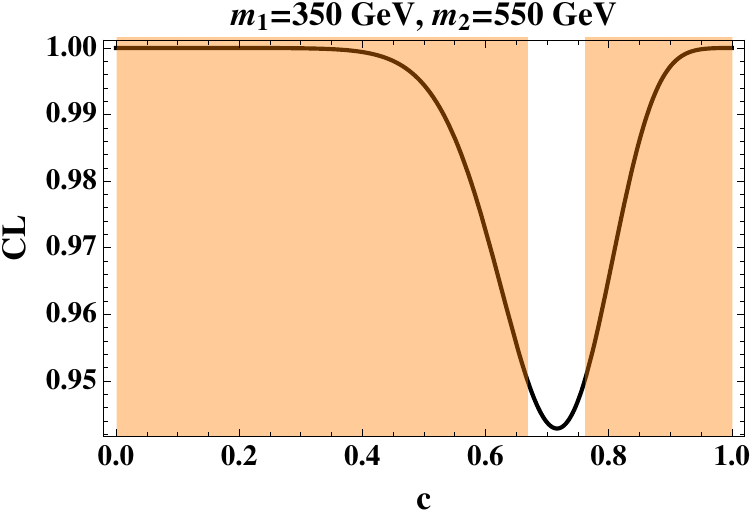}\hfill
\includegraphics[width=0.49\textwidth]{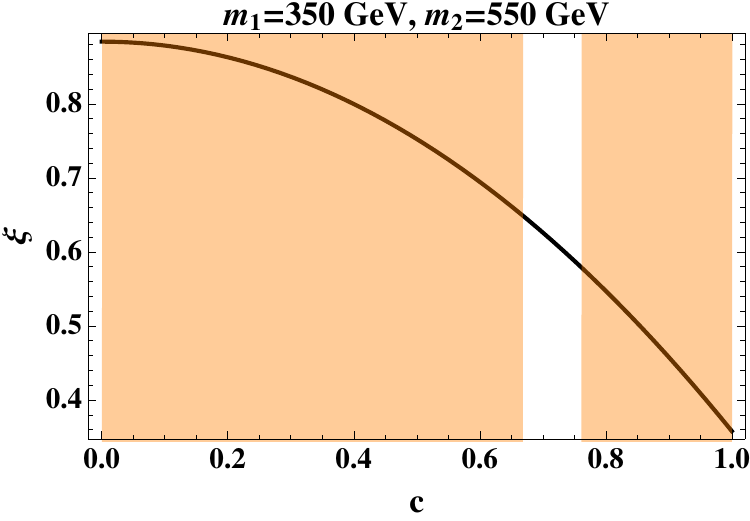}

\caption{\label{fig:Xi_CL_350_450}The confidence level (CL) of exclusion and the tuning parameter $\xi$ as functions of $c=\cos\theta^{ct}_R$ for the case  $m_1=350\gev$ and $m_2=550\gev$. Shaded regions are excluded at 95\%CL.}
\end{figure}

Again, as an example, we pick the mass spectrum $m_1=350\gev$ and $m_2=550\gev$ and study the behaviour of the confidence level and the fine-tuning parameter $\xi$ as a function of the mixing angle, see Fig.\ \ref{fig:Xi_CL_350_450}. The observations are qualitatively similar to the ones we made before in the conservative case. However, now the confidence level increases much more steeply when moving away from the maximal mixing scenario, so that the allowed window for $c$ becomes much smaller.

\begin{figure}
\begin{center}
\includegraphics[width=0.49\textwidth]{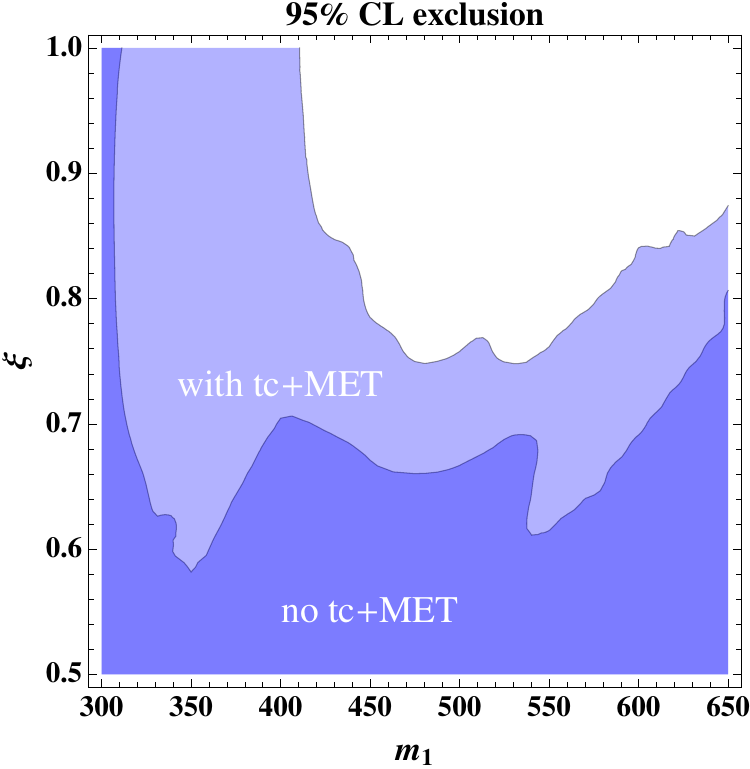}
\end{center}
\caption{\label{fig:Xi_all}The 95\% CL exclusion of the tuning parameter $\xi$ as a function of $m_1$ requiring that $c=\cos\theta^{ct}_R>1/\sqrt2$ so that $\tilde q_1$ is mostly stop-like. The conservative (aggressive) approach corresponds to the lighter (darker) shaded exclusion region.}
\end{figure}

The minimal value of the fine-tuning parameter $\xi$ at fixed value of the stop-like mass $m_1$ is shown in Fig. \ref{fig:Xi_all}. When obtaining the 95\% CL exclusions in the aggressive (dark shaded region) and conservative (light shaded) approaches, the mixing angle was kept above $45^\circ$ so that $\tilde q_1$ is always stop-like. We see that reductions in $\xi$ up to about $40\%$ ($20\%$) are possible in the aggressive (conservative) analysis.
Overall the minimal value of the fine-tuning parameter $\xi$ decreases when going to lower masses $m_1$, and then increases steeply when the experimental bounds are saturated. This saturation is reached for significantly  lower mass values in the aggressive approach (dark blue area) than in the conservative one (light blue area). The two steep minima around $350\gev$ and $550\gev$ obtained in the aggressive approach are again caused by the steeply falling experimental bound on the stop pair production cross section for masses around $380\gev$ (see the black curve in Fig.\ \ref{fig:xsecbounds}).

\subsection{Hunting for the smoking gun}

After estimating the size of the current bounds on the stop-scharm mixing scenario, let us briefly outline how to test the set-up in a dedicated search. Above we have already pointed out the presence of the final state $t\bar c(c\bar t)+\EmissT$, which has previously been discussed in~\cite{Bartl:2010du}.

\begin{figure}
\includegraphics[width=0.49\textwidth]{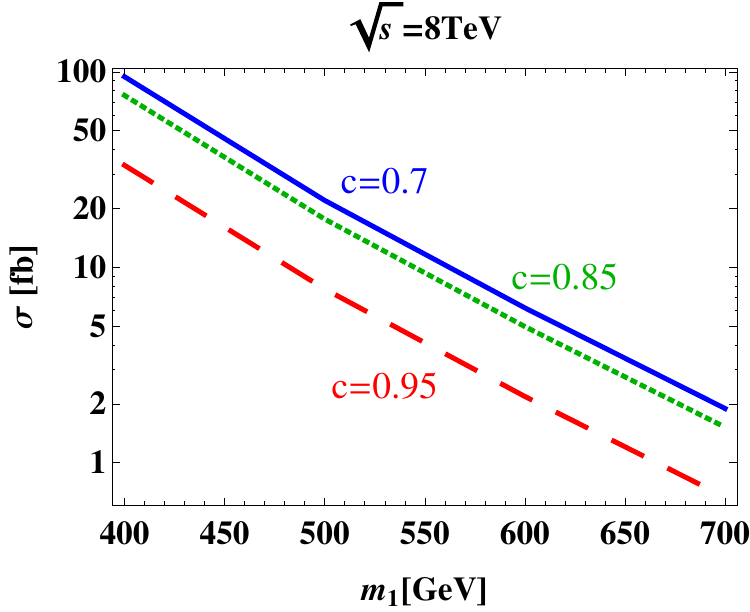}
\hfill
\includegraphics[width=0.49\textwidth]{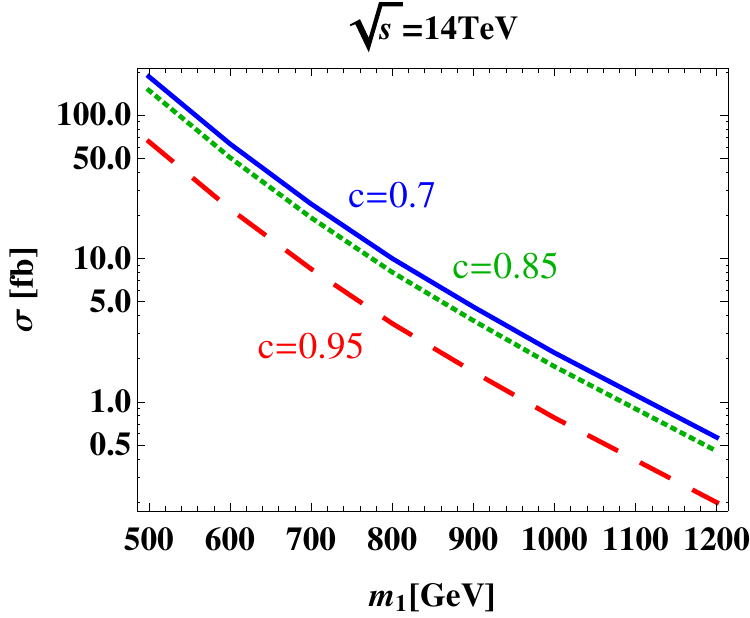}

\caption{\label{fig:tcMET}LO prediction for the $t\bar c(c\bar t)+\EmissT$ signal at the LHC from pair production of the flavour-mixed squark state with mass $m_1$, for ${\sqrt s}=8$~TeV (left panel) and 14~TeV (right panel). Three different choices for the mixing angle $c=\cos\theta^{ct}_R$ are displayed: $c=0.7$ (blue, solid), $c=0.85$ (green, dotted), $c=0.95$ (red, dashed).}
\end{figure}

We use \textsc{MadGraph 5}~\cite{Alwall:2011uj} to simulate the LO prediction for the cross section of $t\bar c(c\bar t)+\EmissT$, stemming from the decay of a mixed $\tilde c_R -\tilde t_R$ state. The result is shown in Fig.\ \ref{fig:tcMET} both for the $8\tev$ and the $14\tev$ runs of the LHC.
We observe that a cross-section of $100\,\text{fb}$ can be reached even at the $8\tev$ run of the LHC, provided a low mass $m_1\sim 400\gev$ and a close to maximal mixing angle. In a dedicated analysis of the data already on tape it should therefore be possible to derive relevant constraints on the $(c,m_1)$ parameter space. 
Needless to say, the $14\tev$ run will be still much more promising. From the right panel of Fig.\ \ref{fig:tcMET} it is clear that cross-sections above $1\,\text{fb}$ can be reached even for masses above $1\tev$.

Since the $t\bar c(c\bar t)+\EmissT$ final state is rather unique, search strategies should ideally involve some mechanism to identify the quark flavours, in addition to {standard $H_T$ and missing energy cuts. While a detailed study is beyond the scope of this paper, we note that the latter cuts generally yield good discriminating power between SUSY signals and the SM background.} 
Depending on the mass spectrum, the top quark might be boosted enough for a top tagging mechanism that makes use of the top jet substructure to become efficient \cite{Kaplan:2008ie,Almeida:2008tp,Ellis:2009me,Plehn:2010st,Rehermann:2010vq,Plehn:2011tf}. In a less boosted scenario mass reconstruction of the hadronically decaying top or the requirement of a $b$-tagged jet and an isolated lepton from a semi-leptonic top are promising.
It remains to be seen whether a charm tagging algorithm can help to extend the reach of the search. Note that charm tagging might be the only way to distinguish the effects of scharm-stop mixing  in this channel from a possible mixing between the right-handed up and top squarks. Only if the gluino is light enough, the up squark pair production receives a significant enhancement from the up quark PDF, resulting in an enhanced cross section of the $t+\text{jet}+\EmissT$ signal.

\subsection{Same-sign tops}

Another possible smoking gun signature of our model would be the observation of same-sign top quarks, a signature basically absent in the SM. 
Generally same-sign squarks can be produced in the process $qq \to \tilde q \tilde q$ which is mediated by a $t$-channel gluino or neutralino. The final state will then be $qq+\EmissT$.
In the absence of flavour violation this process does not lead to the production of same-sign tops since there are no tops in the proton. Consequently observing a same-sign top final state arising from squark pair production would be a clear signal of flavour violation involving the top. In passing we note that same-sign tops could also arise from gluino pair production, where the gluino decays into top-stop. If the decay products of the stop are too soft to pass the $p_T$ cuts, the signal will effectively be $tt+\EmissT$~\cite{Kraml:2005kb}. However at the same time also $\bar t \,\bar t+\EmissT$ (with the same rate) and $ t \bar t+\EmissT$ (with twice the rate) would be produced, making this process distinguishable from the direct same-sign squark production.

\begin{figure}
\centering
\includegraphics[width=0.49\textwidth]{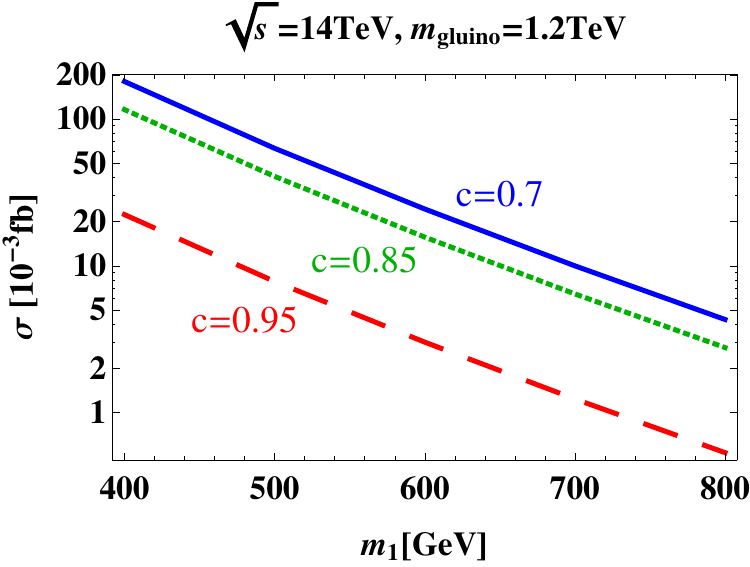}
\caption{\label{fig:samesigntop}LO prediction for the cross-section of same-sign top production at the $14\tev$ LHC run. We assume $m_{\tilde g}=1.2\tev$ and a non-negligible splitting $m_1<m_2$. Three different choices for the mixing angle $c=\cos\theta^{ct}_R$ are displayed: $c=0.7$ (blue, solid), $c=0.85$ (green, dotted), $c=0.95$ (red, dashed).}
\end{figure}

In the presence of stop-scharm mixing same-sign tops can be produced via~\cite{Bartl:2010du} 
\begin{equation}
cc \to \tilde q_i \tilde q_j \to tt+\EmissT\,,
\end{equation}
where $\tilde q_{i,j}$ are the mass eigenstates of the mixed stop-scharm system.
In order to get a notion of the possible size of the effect, we use \textsc{MadGraph 5}~\cite{Alwall:2011uj} to evaluate the LO cross section for same-sign top production at the $14\tev$ run of the LHC. The result is shown in Fig.\ \ref{fig:samesigntop}, where we assumed a gluino mass of $1.2\tev$ and a non-negligible splitting $m_1<m_2$.

It turns out that the production cross section suffers a strong suppression from the smallness of the charm quark PDF and is always below $0.2\,\text{fb}$ even for a light squark around $400\gev$. Taking into account the branching ratio suppression $\sim (2/9)^2$ for semi-leptonic tops leads us to the conclusion that stop-scharm mixing cannot be the origin of a possible observation of same-sign tops at the $14\tev$ run of the LHC. 
At the luminosity upgraded SLHC an observation should be possible in the most favourable case of low masses and a large mixing angle.

\begin{figure}
\centering
\includegraphics[width=0.49\textwidth]{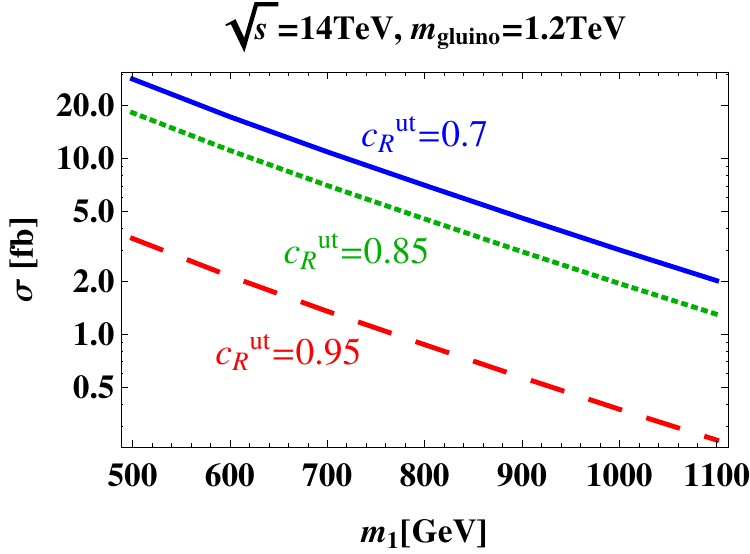}
\caption{\label{fig:samesigntop-13}LO prediction for the cross-section of same-sign top production at the $14\tev$ LHC run, arising from a large $\tilde t_R - \tilde u_R$ mixing. We assume $m_{\tilde g}=1.2\tev$ and a non-negligible splitting $m_1<m_2$. Three different choices for the mixing angle $c_R^{ut} = \cos\theta^{ut}_R$ are displayed: $c_R^{ut}=0.7$ (blue, solid), $c_R^{ut}=0.85$ (green, dotted), $c_R^{ut}=0.95$ (red, dashed).}
\end{figure}

If on the other hand the same-sign top signature is generated through large up-top squark mixing, we can expect it to be accessible in the $14\tev$ run. We show the corresponding LO cross section in Fig.\ \ref{fig:samesigntop-13}. For identical masses and mixing the cross section is more than two orders of magnitude larger than in the case of scharm-stop mixing, thanks to the large up quark content of the proton. Additionally in this case, contrary to the stop-scharm mixing scenario, the cross section for $tt+\EmissT$ is significantly larger than for $\bar t \bar t +\EmissT$.  The observation or non-observation of the same-sign top signal in the $14\tev$ run is therefore an important tool to distinguish whether large stop-sup mixing or stop-scharm mixing is realised in nature.
Note however that due to the PDF enhancement the squark masses in this case are much more strongly constrained~\cite{Mahbubani:2012qq}.

\section{Direct CP violation in charm decays}\label{sec:charmCPV}

The presence of a large mixing between the right-handed charm and top squarks may also have
interesting implications in low-energy processes involving up-quarks.
In particular, the direct charm CP violation in $D$ meson systems $D\to K^+K^-/\pi^+\pi^-$ which has  recently been measured
$\Delta a_{CP}=a^{\rm dir}_K-a^{\rm dir}_\pi=-(0.65\pm 0.18)\%$~\cite{lhcb,Aaltonen:2011se, Staric:2008rx, Aubert:2007if, HFAG}, can be explained in supersymmetry~\cite{Grossman:2006jg,Giudice:2012qq}.
In our setup,
we can effectively generate the necessary $c\to u$ transition through the combination
$(c_{ L}^{ ut} s_{ L}^{ ut})
(m_t A_t/\tilde m_q^2)
(c_{ R}^{ ct} s_{R}^{ct})$~\cite{Giudice:2012qq}. Here we define $c_{L,R}^{ij}=\cos\theta_{L,R}^{ij}$ and $s_{L,R}^{ij}=\sin\theta_{L,R}^{ij}$. Therefore,
assuming maximal CP violation, we find that
\be
\left|\Delta a^{\rm SUSY}_{CP} \right| \approx 0.6\%~
\bigg|\,
\left(\frac{c^{ut}_{ L} s^{ ut}_{ L}}{\lambda^3}\right)
\left(\frac{m_t A_t/\tilde m_q^2}{0.1}\right)
\left(\frac{c^{ ct}_{ R} s^{ ct}_{ R}}{\mathcal{O}(1)}\right)
\bigg|
\left(\frac{{\rm TeV}}{\tilde m}\right)~.
\label{eq:acp1}
\ee
Notice that the off-diagonal elements of the CKM matrix provide 
reference values for the mixing in the left-handed sector, as for instance
in the case of down-type alignment and split families where we expect
$c_{ L}^{ut} s^{ut}_{L} \approx
V_{ub} ~{\Delta\tilde m^2_{31}\over\tilde m^2}$
with $\Delta\tilde m^2_{31}$ being the square mass splitting between the
first and third generations of left-handed squarks.
Moreover, within the minimal supersymmetric model, the lightest Higgs boson mass $m_h \simeq 125$~GeV
 favours a large trilinear coupling $A_t$.

Therefore, whenever the right-handed stop-scharm mixing angle is large, $c^{ct}_{R} s_{R}^{ct}\sim\mathcal{O}(1)$,
and the squark and gluino masses are around $1~$TeV, it is easy to generate a direct
charm CP violation at the percent level in large regions of the supersymmetric parameter space.

\section{Conclusions}

The flavour composition of squarks can significantly affect the squark searches at the LHC. Large mixings between right-handed top and charm squarks are left unconstrained by low-energy flavour processes (in the limit in which the first generation of up squarks is aligned with quarks) and can appreciably lower the bounds on the stop-like mass. Depending on the mixing angle the bound on the stop-like mass gets relaxed by about 100\,GeV or more compared to  the case of no flavour mixing. As a result, and somewhat surprisingly, a nontrivial flavour structure in the squark sector can lead to a mild reduction of fine tuning. 
Once more data becomes available, the difference between the bounds on the stop and scharm masses
will only become larger, thus further increasing the relative gain in naturalness in the presence of mixing.

Large stop-scharm mixing can lead to interesting experimental signatures at the LHC. The smoking gun is the presence of the signal in the $t\bar c(c\bar t)+\EmissT$ channel that can be immediately searched for. Another interesting signature is same-sign top production from the production chain $qq\to \tilde q_i\tilde q_j\to tt+\EmissT$. The search for this signature is possible even in the most optimistic scenario only on longer time scales after the LHC upgrade.

The large stop-scharm mixing, if accompanied by a sizeable soft $A$-term for the stop sector, might also lead to new CP-violating effects in $D\to K^+K^-/\pi^+\pi^-$ as recently observed by LHCb. While data can in principle be accounted for within the Standard Model~\cite{Brod:2011re}, its close connection to high-$p_T$ observables  makes the stop-scharm mixing particularly interesting.

Let us conclude by stressing that the implications of stop-scharm mixing studied in the present paper hold well beyond the minimal supersymmetric scenario. The phenomenology of squark flavour mixing and their decay into quarks are common to a large class of supersymmetric models.
Hence the beneficial effects of $\tilde t_R - \tilde c_R$ mixing on fine-tuning together with an interesting phenomenology provide a strong motivation to explore this possibility experimentally. 
\vspace{5mm}

{\bf Acknowledgements:} GP thanks Josh Ruderman and Raman Sundrum for  useful discussions. 
GP is supported by grants from GIF, IRG, ISF and Minerva. JZ was supported in part by the U.S. National Science Foundation under CAREER Grant PHY-1151392.

\end{document}